\documentclass{llncs}

\usepackage[utf8]{inputenc}
\usepackage[english]{babel}
\usepackage[T1]{fontenc}
\usepackage{microtype}
\usepackage{url}

\usepackage{amsmath}
\usepackage[bold,small,langfont=roman]{complexity}
\newclass{\SharpP}{\boldsymbol{\#}P}

\title{Unconventional Complexity Classes\\in Unconventional Computing}
\subtitle{Extended Abstract}

\author{Antonio E. Porreca\orcidID{0000-0003-1544-028X}}

\institute{
   Aix-Marseille Université, CNRS, LIS, Marseille, France\\
   \email{antonio.porreca@lis-lab.fr}
}

\begin{document}

\maketitle

\begin{abstract}
Many unconventional computing models, including some that appear to be quite different from traditional ones such as Turing machines, happen to characterise either the complexity class~$\P$ or~$\PSPACE$ when working in deterministic polynomial time (and in the maximally parallel way, where this applies). We discuss variants of cellular automata and membrane systems that escape this dichotomy and characterise intermediate complexity classes, usually defined in terms of Turing machines with oracles, as well as some possible reasons why this happens.
\end{abstract}

\section{Introduction}

Unconventional computing models, particularly biologically inspired ones, appeared very early in the history of modern computer science, notably with cellular automata. Once new a model has been introduced, it is natural to analyse whether it is universal, in the sense that it can compute anything a Turing machine can; this happens quite often, since very little is actually needed in order to achieve universality (for instance, it usually suffices to simulate incrementation and conditional jump instructions~\cite{Minsky1967a}). These models are thus unconventional in terms of mechanism rather than in terms of computing power.

Nonetheless, a ``conventional'' computing power can be achieved with unconventional efficiency. This is where computational complexity questions become relevant. Some unconventional models, mostly sequential ones or parallel ones with a polynomial number of ``processors'', can simulate and be simulated by deterministic Turing machines with a polynomial-time overhead, and thus in particular they characterise the complexity class~$\P$ when working in polynomial time. Emde Boas refers to those models as the \emph{first machine class}~\cite{VanEmdeBoas1990a}. This includes random-access machines with constant-time addition and subtraction~\cite{Papadimitriou1993a} and, as detailed below, traditional cellular automata.

Other models, with less restrictions on parallelism or with more powerful elementary operations, characterise in polynomial time what deterministic Turing machines compute in polynomial \emph{space}. This is called the \emph{second machine class}~\cite{VanEmdeBoas1990a} and includes parallel models such as tree-shaped hierarchies of processes generated by the \texttt{fork} system call under Unix (and running on an unbounded number of processors), as well as sequential ones such as random-access machines with constant-time multiplication and division~\cite{Bertoni1981a}, and nondeterministic one such as alternating Turing machines~\cite{Papadimitriou1993a}.

Not all unconventional (or conventional, for that matter) computing models fall either into the first or the second machine class, but it happens often enough that one might be interested in investigating what features cause this behaviour, and what can be added, for instance, to a sequential model of the first class in order to solve more problems efficiently without always obtaining~$\PSPACE$\footnote{Of course, here we work under the hypothesis that~$\P \ne \PSPACE$ and that the intermediate complexity classes mentioned later are also distinct from both of them.}. One way to do that is to introduce randomisation~\cite[Chapter~11]{Papadimitriou1993a} or quantum computing features~\cite{Watrous2009a}, but we will explicitly exclude these from the present discussion, both because they are outside the area of expertise of the author, and because they require an alternative definition of output which takes ``wrong guesses'' into account. Rather, let us focus on deterministic models and consider two examples, from the theory of cellular automata and from membrane computing.

\section{Cellular Automata}

One-dimensional cellular automata can simulate Turing machines with a poly\-nomial-time slowdown, trivially by having a large enough set of states (or, equivalently, neighbourhood size) and storing the symbols, state, and tape head position of the Turing machine being simulated as states of the automaton\footnote{With a more sophisticated reasoning, one can prove that the rule 110 automaton can also efficiently simulate Turing machines~\cite{Neary2006a}.}~\cite{Smith1972a}. This is, however, a strictly sequential simulation carried out by a parallel model, and one might wonder whether cellular automata can be significantly more efficient when that parallelism is actually exploited. The answer is, however, readily seen to be negative: for any integer~$d$, a $d$-dimensional cellular automata starting from a finite (non-quiescent) initial configuration of diameter~$2r$ can always be simulated sequentially in polynomial time with respect to~$r$. This happen because the volume of the smallest sphere containing the non-quiescent portion of the configuration is polynomial with respect to the radius, specifically~$O(r^d)$, and the radius can only increase by one at each computation step of the automaton. Standard $d$-dimensional cellular automata are thus first class machines, despite their parallelism.

In order to solve harder problems we must then switch to cellular automata over non-hypercubic grids. The classic example from the literature is given by hyperbolic cellular automata~\cite{Margenstern2017a}. Switching from Euclidean to hyperbolic geometry allows us to construct regular pentagonal grids, and the number of cells contained in a sphere of radius~$r$ is not polynomial anymore, but rather \emph{exponential}. This allows many more cells to be active in parallel at any given time. More specifically, an infinite binary tree, whose branches represent communication channels, can be embedded in the pentagonal grid and exploited in order to solve~$\PSPACE$-complete problems in polynomial time. For instance, consider the quantified $3\SAT$ problem for a formula of~$n$ variables: the idea is to explore the~$2^n$ possible truth assignments along~$2^n$ distinct paths in the tree, evaluating the formula under each assignment separately, then propagating the results back towards the root; during this backpropagation phase, the truth values in two adjacent paths are combined by conjunction or disjunction, depending on the alternation of quantifiers in the formula, and the final result is obtained at the root. This is essentially a simulation of the computation tree of an alternating Turing machine. As a consequence, a hyperbolic computation space brings cellular automata to the second machine class.

An interesting variant of cellular automata, introduced by Modanese and Worsch~\cite{Modanese2016a}, falls between~$\P$ and~$\PSPACE$ in polynomial time. These are \emph{shrinking and expanding cellular automata}, where a cell can not only update its state based on its neighbourhood, but also delete itself (shrinking); furthermore, new cells can be created between two existing ones (expanding). While the full model once again characterises~$\PSPACE$ by simulating the computation trees of alternating Turing machines~\cite{Modanese2016a}, disallowing shrinking decreases the efficiency of the model. Essentially, in this model the information can only be propagated towards the leaf of the simulated computation tree, but since the distance between leaves become exponential in time, the information cannot be propagated back towards the root, and an unanimity acceptance condition is used instead~\cite{Modanese2022a}. As a result, expanding cellular automata characterise in polynomial time the class of problems \emph{truth-table reducible to~$\NP$ problems}. This is conjecturally weaker than~$\PSPACE$, but large enough to include both~$\NP$ and~$\coNP$ and, as such, fits our intuition of \emph{unconventional complexity class}.

\section{Membrane Computing}

Membrane systems~\cite{Paun2000a}, also called P~systems, are models inspired by the internal structure and functioning of biological cells. In their basic models, they consist of a tree-shaped hierarchy of nested membranes containing a \emph{multiset} of molecules. The systems evolves inside each membrane by applying multiset-rewriting rules inspired by biochemical reactions; furthermore, membranes are selectively permeable, and can send molecules to, or receive them from adjacent regions. Finally, membranes can \emph{divide} by fission, and their content is duplicated (with the exception of the triggering molecules, which differentiates the resulting membranes). These computation rules are applied, by default, in the \emph{maximally parallel way}: if a molecule or membrane \emph{can} be subject to at least one rule, then it \emph{must} do so\footnote{Strictly speaking, if multiple rules are applicable, then one is chosen nondeterministically, but a \emph{confluence} condition is usually imposed when solving decision problems (all computations must give the same result in the end). In all interesting cases known to the author, the systems can actually be made deterministic, although a general proof of this statement is still missing.}.

When no membrane division occurs, but only chemical reactions, the resulting systems belong to first machine class~\cite{Zandron2001a,Alhazov2013a}. However, when membranes can divide at arbitrary nesting depths, thus duplicating whole nested membrane substructures, exponential-size tree-shaped structures can be created in polynomial time, which can be exploited in order to simulate alternating Turing machines and prove, once again, an exact characterisation of~$\PSPACE$ in polynomial time~\cite{Sosik2007a}.

By restricting communication in only one direction (from the inside to the outside, i.e., membranes can send out molecules but never absorb them), similarly to the restriction on expanding cellular automata, the efficiency of membrane systems decreases and we obtain a characterisation of~$\P^\NP$ in polynomial time~\cite{Leporati2016b}. This is the class of problems solved in polynomial time by deterministic Turing machines with access to an oracle for an~$\NP$ problem; this can also be viewed as the class of problems Cook-reducible to~$\NP$, which highlights the similarity with expanding cellular automata even more. Another unconventional complexity class!

A different constraint we can impose is not related to the direction of communication but rather on the depth of the tree we can build. If communication is bidirectional, but the membrane systems are required to be \emph{shallow} (only one level of nested membranes) or, more generally, if membranes can only divide if they do not contain further membranes, then we obtain a characterisation~\cite{Leporati2014d} of the complexity class~$\P^\SharpP$; this is the class of problems solvable in polynomial time with access to an oracle for a counting problem~\cite{Papadimitriou1993a}, or equivalently those Cook-reducible to~$\SharpP$.

\section{Conclusions}

The examples we described highlight how geometric (Euclidean vs hyperbolic space) and communication (monodirectional vs bidirectional) constraints can affect the efficiency of parallel unconventional computing models. Can these consideration be abstracted and formalised in a more general framework, or at least generalised to other communication topologies or geometries and for other computation models?

Another interesting aspect of the results on expanding cellular automata and membrane systems is that they provide a characterisation of complexity classes such as~$\P^\NP$ and~$\P^\SharpP$ in terms of a \emph{concrete, deterministic} model of computation; by this, we do not necessarily mean a \emph{realistic} model (whenever exponentially many processors are needed, the shape of the approximately Euclidean space we live in is always a bottleneck~\cite{Leporati2015c}) but a deterministic model whose instantaneous configurations can be described in full detail, quite unlike the original definition of these complexity classes in terms of black-box oracles. Can other ``abstract'' complexity classes like these be captured by more concrete computing models?

\subsection*{Acknowledgements}
This work was partially supported by the EU project MSCA-SE-101131549 ``ACANCOS''.

\bibliographystyle{splncs04}
\bibliography{Bibliography}

\end{document}